\documentclass
[prc,twocolumn,showpacs,preprintnumbers,amsmath,amssymb,superscriptaddress,floatfix,nofootinbib]{revtex4}
\usepackage{graphicx}
\usepackage{amsmath}
\usepackage{amsfonts}
\usepackage{slashed}
\usepackage{amssymb}
\newcommand{\smsm}[1]{\scriptscriptstyle{\scriptscriptstyle{#1}}}

  \def\CL{{\cal L}}
\def\CM{{\cal M}}

\begin{document}

\title{Polarization of $\Lambda(1405)$ in the $\gamma p \rightarrow K^+ \pi \Sigma$ reaction}

\author{Ke Wang}\author{Bo-Chao Liu} \email{liubc@xjtu.edu.cn} \affiliation{School of Physics, Xi'an
Jiaotong University, Xi'an, Shannxi 710049, China}

\begin{abstract}

In this paper, we study the polarization of the $\Lambda(1405)$ in
the $\gamma p \rightarrow K^+ \pi \Sigma$ reaction within an
effective Lagrangian approach and isobar model. In our model, the
$\Lambda(1405)$ is excited through the t-channel $K/K^*$ exchanges
and u-channel hyperon exchange. Compared to previous studies, we
also include the contribution from a contact term, which is
necessary for our model to interpret the polarization of the
$\Lambda(1405)$. In addition, we also discuss the possibility to
verify the proposed two-pole structure of the $\Lambda(1405)$ using
the polarization data. We find that the polarization of the
$\Lambda(1405)$ or the polarization of the final $\Sigma$ in this
reaction is sensitive to the invariant mass $M_{\pi\Sigma}$. Thus
the measurement of the dependence of the $\Lambda(1405)$
polarization on the $M_{\pi\Sigma}$ can offer valuable information
about the pole structure of the $\Lambda(1405)$.

\end{abstract}
\maketitle

\section*{INTRODUCTION}
The structure and properties of the $\Lambda(1405)$ is an
interesting and important topic in hadron physics, which has
attracted a lot of interest since its existence was
predicted\cite{Dalitz:1959dn,Dalitz:1960du}. Due to the attractive
interaction between antikaons and nucleons, the $\Lambda(1405)$ may
be a quasibound molecular state of the $\bar KN$ system. While, in
the conventional quark model the $\Lambda(1405)$ was also described
as a $p$-wave state of three-quark system\cite{Isgur:1978xj}. In the
1990's, the $\Lambda(1405)$ was investigated within the Chiral
Unitary approach, and it was found that the $\Lambda(1405)$ could be
dynamically generated, i.e. appearing as poles in the amplitude,
through the SU(3) dynamics\cite{Fink:1989uk}. An interesting finding
in this approach is that, in contrast to the conventional opinions,
the observed bump of the $\Lambda(1405)$ is in fact due to two poles
in the amplitude. This finding was confirmed by a series of further
theoretical stuides\cite{Oller:2000fj,Oset:2001cn,Jido:2003cb}. The
possibility of existing of two poles in the $\Lambda(1405)$ region
has stimulated a lot of further efforts to explore the nature of the
$\Lambda(1405)$. Unfortunately, up to now there is still no final
conclusion about whether the two-pole structure exists or not. It is
fair to say that we still do not understand the nature of the
$\Lambda(1405)$ very well.

Among the various studies on the $\Lambda(1405)$, we are interested
in the $\Lambda(1405)$ production in the photo induced process. In
Refs.\cite{Moriya:2013hwg,Moriya:2013eb}, the study of the
$\Lambda(1405)$ production in the $\gamma p\to K \pi\Sigma$ reaction
was reported by the CLAS Collaboration. They measured the angular
distribution of the final $K^+$ and the invariant mass spectrum of
$\pi\Sigma$, which offer a good opportunity for studying the
properties of the $\Lambda(1405)$ and verifying its possible
two-pole structure. These data were analyzed in
Refs.\cite{Nakamura:2013boa,Roca:2013av,Nam:2015yoa,Oset:2015ksa}.
However, it seems current data cannot offer enough constraints on
the model. Therefore, it is still not possible to draw the
conclusion about the two-pole conjecture. Besides the measurement of
the angular distribution and the invariant mass spectrum, some
progress was also made in identifying the quantum numbers of this
resonance. In Ref.\cite{Moriya:2014kpv}, the CLAS Collaboration
reported their results on determining the quantum numbers of the
$\Lambda(1405)$, which confirmed the quantum numbers $J^P$ of the
$\Lambda(1405)$ is $\frac{1}{2}^-$. What makes this experiment
interesting for us is the idea of the measurement of the
$\Lambda(1405)$ polarization in their work. In order to measure the
spin and parity of the $\Lambda(1405)$, the $\Lambda(1405)$ is
assumed to be produced polarized in this reaction. Even though the
clarifying of the mechanism for the $\Lambda(1405)$ polarization in
this reaction is not necessary for the purpose of measuring its
quantum numbers, it is certainly interesting and important for
understanding the reaction mechanisms. Since in the experimental
analysis\cite{Moriya:2014kpv} the $\Lambda(1405)$ is treated as a
single resonance, it will also be interesting to discuss the
possible effects if considering the two-pole structure of the
$\Lambda(1405)$. In previous studies, these issues were not
considered. So the main goal of the present work is twofold. First,
we hope to discuss the mechanism for the polarization of the
$\Lambda(1405)$ in this reaction. In fact, we find the polarization
data can offer further constraints on the model, which are helpful
for understanding the reaction mechanisms. Second, we hope to
discuss the effects due to the pole structure of the $\Lambda(1405)$
on the $\Lambda(1405)$ polarization. This issue is interesting
because it may offer a new way to verify the two-pole picture of the
$\Lambda(1405)$. Since we hope to concentrate on the mechanism for
the $\Lambda(1405)$ polarization in the present work, we will only
consider the experimental data at the center-of-mass energies
ranging from 2.3 GeV to 2.8 GeV, where the s-channel nucleon
resonance contribution is small\cite{Kim:2017nxg} and the
polarization of the $\Lambda(1405)$ was measured.

This paper is organized as follows. In Sec. II, the theoretical
framework and ingredients are presented. In Sec. III, the numerical
results are presented with some discussions. Finally, the paper ends
with a short summary in Sec. IV.
\section*{MODEL AND INGREDIENTS}

\begin{figure}[htbp]
\begin{center}
\includegraphics[scale=0.45]{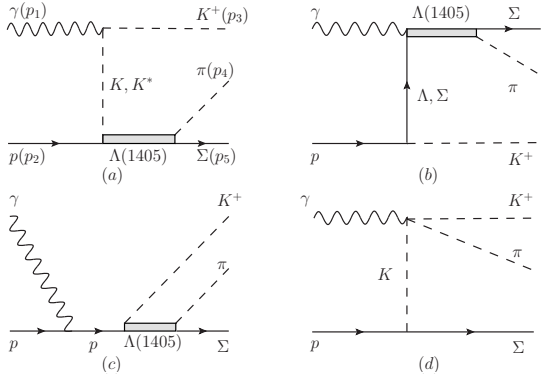}
\caption{Feynman diagrams for the $\gamma p \rightarrow K^+ \pi
\Sigma$ reaction.} \label{feynman}
\end{center}
\end{figure}
In the present work, we study the $\gamma p \rightarrow K^+ \pi
\Sigma$ reaction at the c.m. energies ranging from $2.3$ GeV to
$2.8$ GeV within an effective Lagrangian approach and isobar model.
As mentioned in Sec. I, at these energies the $s$-channel nucleon
resonance contribution can be ignored. To describe this reaction, we
consider the Feynman diagrams shown in Fig.\ref{feynman}, which
include the pseudoscalar and vector meson exchanges in the
$t$-channel, the hyperon exchange in the $u$-channel, nucleon pole
term in the $s$-channel and a contact term. To evaluate these
Feynman diagrams, the effective Lagrangian densities for the
interaction vertices are needed. For the electromagnetic interaction
vertices, we have the interaction Lagrangian
densities\cite{Kim:2017nxg,Vysotsky:2015mks}:
\begin{eqnarray}
    \CL_{\gamma K K} &=&
-ie_K [K^\dag(\partial_\mu K)-(\partial_\mu K^\dag)K]A^\mu , \\
    \CL_{\gamma K K^*} &=&
g_{\gamma K K^*} \varepsilon^{\mu \nu \alpha \beta} \partial_\mu A_\nu \nonumber \\ &&
[(\partial_\alpha K^{*-}_\beta)K^+ + K^-(\partial_\alpha K^{*+}_\beta)] , \\
    \CL_{\gamma N N} &=&
-{\bar N}\Big[e_N\gamma_\mu-\frac{e\kappa_N}{2M_N}\sigma_{\mu \nu}\partial^\nu \Big] A^\mu N , \\
    \CL_{\gamma Y \Lambda^*} &=&
\frac{e\mu_{\smsm{\Lambda^*Y}}}{2M_N}{\bar Y} \gamma_5 \sigma_{\mu \nu} \partial^\nu A^\mu \Lambda^* +\mathrm{H.c.} ,\\
    \CL_{C} &=&
-\frac{ieg_c}{8 \pi^2 F^3_\pi} \varepsilon^{\mu \nu \alpha \beta}
F_{\mu \nu} \pi \partial_\alpha K \partial_\beta {\bar K}  ,
\end{eqnarray}
where $\varLambda^*, A_\mu, Y, K$ and $K^*$ denote the
$\Lambda(1405)$ , photon, hyperon($\Sigma$ or $\Lambda$), $K$ and
$K^*$ fields, respectively. The charge of electron $e$ and $\pi$
decay constant $F_\pi$ are taken as the usual values, i.e.
$e=\sqrt{4\pi/137}$ and $F_\pi=92.2$MeV\cite{Vysotsky:2015mks}.
$g_c$ describes the coupling strength of the contact term, and we
treat it as a free parameter in this work. Other coupling constants
in the Lagrangians are taken from previous studies, which are
determined by either fitting the experimental data or theoretical
predictions. The values for these parameters and relevant references
are listed in Table \ref{tab1}.

For the strong interaction vertices, the interaction Lagrangian
densities can be written as \cite{Kim:2017nxg,Nam:2015yoa}:
\begin{eqnarray}
    \CL_{K N Y} &=& -ig_{K N Y} {\bar N} \gamma_5 Y K +\mathrm{H.c.} , \\
     \CL_{\Lambda^* K N} &=& -ig_{\Lambda^* K N} {\bar N} \Lambda^* K +\mathrm{H.c.} , \\
    \CL_{\Lambda^* K^* N} &=& -g_{\Lambda^* K^* N} {\bar N} \gamma_5 \gamma_\mu \Lambda^* K^{*\mu} +\mathrm{H.c.}, \\
    \CL_{\Lambda^* \pi \Sigma} &=& ig_{\Lambda^* \pi \Sigma} {\bar \Lambda^*} {\vec \pi} \cdot {\vec
    \Sigma}+\mathrm{H.c.}.
\end{eqnarray}
Here we take the coupling constants $g_{KN\Lambda}$ and
$g_{KN\Sigma}$ from the Nijmegen soft-core
potential\cite{Stoks:1999bz}. Current knowledge of the
 $g_{\Lambda^* K^* N}$ is still rather limited,
so we take the value from Ref.\cite{Khemchandani:2011mf}, where the
value of $g_{\Lambda^*K^*N}$ is obtained by averaging the
predictions of various chiral-unitary models(ChUM).

To take into account the internal structure of hadrons, we have
introduced form factors in the calculations. In this work, the form
factor for intermediate hadrons is taken as \cite{Kim:2017nxg}
\begin{eqnarray}
    F(q,m) &=& \bigg ( \frac{\Lambda^4}{\Lambda^4+(q^2-m^2)^2} \bigg )^2,
\end{eqnarray}
where $q$ and $m$ are the momentum and mass of the exchanged
particles. We take $\Lambda_M=2.0$ GeV for meson
exchange\cite{Nam:2015yoa}, and fit the $\Lambda_B$, i.e. the cutoff
parameter for baryon exchange, to the experiment data. The
propagators for various particles are adopted as below:
\begin{eqnarray}
    G_K(q) &=& \frac{i}{q^2-m^2}
\end{eqnarray}
for $K$,
\begin{eqnarray}
    G^{\mu \nu}_{K^*}(q) &=& -\frac{i(g^{\mu \nu}-q^\mu q^\nu /q^2)}{q^2-m^2}
\end{eqnarray}
for $K^*$,\begin{eqnarray}
    G_{N/Y}(q) &=& \frac{i(\slashed{q}+m)}{q^2-m^2}
\end{eqnarray} for the nucleon or hyperon and
\begin{eqnarray}
    G_{\Lambda^*}(q) &=& \frac{i(\slashed{q}+m)}{q^2-m^2+im\Gamma(q^2)}
\end{eqnarray}
for the $\Lambda(1405)$, where $q$, $m$ and $\Gamma(q^2)$ are the
four-momentum, mass and width of the exchanged particles
respectively. Here we use an energy-dependent form for the width of
$\Lambda(1405)$. The energy-dependent form is taken as
follows \cite{Xie:2013wfa}:
\begin{eqnarray}
    \Gamma(q^2) &=& \Gamma_{\pi\Sigma}(q^2)+\Gamma_{\bar KN}(q^2), \\
    \Gamma_{\pi\Sigma}(q^2) &=& \frac{3g_{\Lambda^* \pi \Sigma}^2 |\vec p_{\pi\Sigma}(q^2)|} {4 \pi
    \sqrt{q^2}} \nonumber
    \\ && \times \bigg(M_{\Sigma} + \sqrt{M_{\Sigma}^2+|\vec p_{\pi\Sigma}(q^2)|^2} \bigg), \\
    \Gamma_{\bar KN}(q^2) &=& \frac{2g_{\Lambda^* \bar K N}^2 |\vec p_{\bar KN}(q^2)|} {4 \pi \sqrt{q^2}}
    \theta(\sqrt{q^2}-M_{\bar K}-M_N) \nonumber
    \\ && \times \bigg( M_N + \sqrt{M_N^2+|\vec p_{\bar KN}(q^2)|^2} \bigg),
\end{eqnarray}
where $ \vec p_{\pi\Sigma} $ or $ \vec p_{\bar KN} $ denotes the
momentum of final particles in the rest frame of $\Lambda^*$. Below
the $\bar KN$ threshold, the momentum $\vec {p}_{\bar KN}$ is taken
as zero.

With the ingredients presented above, the amplitudes for the $\gamma
p \rightarrow K^+\Lambda^*(\to \pi \Sigma)$ process can be obtained
in a standard way, and we get

\begin{eqnarray}
    {\cal M}_K &=& 2eg_{\Lambda^* K N}g_{\Lambda^* \pi \Sigma} \bar u(p_5,\lambda_\Sigma) G_{\Lambda^*}(q_{\Lambda^*}) \nonumber \\ && p_3 \cdot \epsilon u(p_2,\lambda_p) G_K(q_K) F_{KN},
    \\
    {\cal M}_N &=& eg_{\Lambda^* K N}g_{\Lambda^* \pi \Sigma} \bar u(p_5,\lambda_\Sigma) G_{\Lambda^*}(q_{\Lambda^*}) G_{N}(q_N) \nonumber \\ && \Big(\gamma^\mu + \frac{i\kappa_N}{2M_N} \sigma^{\mu \nu} p_{1 \nu}\Big) \epsilon_\mu u(p_2,\lambda_p) F_{KN},
    \\
    {\cal M}_{K^*} &=& ig_{\gamma K K^*}g_{\Lambda^* K^* N}g_{\Lambda^* \pi \Sigma} \bar u(p_5,\lambda_\Sigma)
    G_{\Lambda^*}(q_{\Lambda^*}) \varepsilon^{\mu \nu \alpha \beta} \nonumber \\ && p_{1\mu} \epsilon_\nu q_{K^*\alpha} G^{K^*}_{\beta \rho}(q_{K^*}) \gamma_5 \gamma^\rho u(p_2,\lambda_p) F_{K^*},
    \\
    {\cal M}_Y &=& \frac{ie\mu_{\smsm{\Lambda^*Y}} g_{K N Y}g_{\Lambda^* \pi \Sigma}}{2M_N}
    \bar u(p_5,\lambda_\Sigma) G_{\Lambda^*}(q_{\Lambda^*}) \nonumber \\ && \gamma_5 \sigma_{\mu \nu} \epsilon^\mu p^{1\nu} G_{Y}(q_Y) \gamma_5 u(p_2,\lambda_p) F_{Y},
    \\
    {\cal M}_{C} &=& g_c \frac{eg_{K N \Sigma}}{4\pi^2F^3_\pi} \bar u(p_5,\lambda_\Sigma) \gamma_5 u(p_2,\lambda_p) \nonumber \\ &&
    \varepsilon^{\mu \nu \alpha \beta} p_{1\mu} \epsilon_\nu q_{K\alpha} p_{3\beta} G_K(q_{K})  F_{K},
\end{eqnarray}
where $p_i$ represents the four-momentum of the particles as denoted
in Fig.\ref{feynman} and $Y$ denotes $\Lambda$ or $\Sigma$.
$F_{M/B}$ is the form factor considered for meson or baryon. To
restore the gauge invariance of the amplitude, we have adopted the
approach in Ref.\cite{Kim:2017nxg} and defined the common form
factor as
\begin{eqnarray}
    F_{KN} &=& F_K + F_N - F_K F_N.
\end{eqnarray}

In previous studies\cite{Kim:2017nxg}, it was shown that in the
energy region under study the Regge approach was successful in
describing the reaction. Following their works, we also adopt the
Regge approach in the present model. To do that, we need to replace
the $t$-channel meson propagators in the amplitudes with the Regge
propagators \cite{Kim:2017nxg,Wang:2019uwk}:
\begin{eqnarray}
\frac{1}{t-M^2_K} &\rightarrow& \Big ( \frac{s}{s_0} \Big
)^{\alpha_K}\frac{\pi \alpha'_K}{sin(\pi \alpha_K)} \frac{1}{\Gamma(1+\alpha_K)}, \nonumber \\
\frac{1}{t-M^2_{K^*}} &\rightarrow& \Big ( \frac{s}{s_0} \Big
)^{\alpha_{K^*}-1} \frac{\pi \alpha'_{K^*}}{sin(\pi \alpha_{K^*})}
\frac{1}{\Gamma(\alpha_{K^*})},
\end{eqnarray}
where $t$ denotes the Mandelstam variable, and Regge trajectories
read~\cite{Kim:2017nxg}
\begin{eqnarray}
    \alpha_K &=& \alpha_K(t) = \frac{0.7}{GeV^2} (t-M^2_K), \nonumber \\
    \alpha_{K^*} &=& \alpha_{K^*}(t) = \frac{0.83}{GeV^2} t + 0.25.
\end{eqnarray}
The slope parameter is defined as $\alpha'_{K,K^*} \equiv \partial
\alpha_{K,K^*}(t)/\partial t$, and the energy scale parameter $s_0$
is chosen to be $1$ GeV$^2$\cite{Kim:2017nxg,Wang:2019uwk}.

The total amplitude $\CM$ is obtained by the summation of the
individual amplitudes. The differential and total cross sections for
this reaction then can be calculated through
\begin{eqnarray}
d\sigma &=& \frac{1}{8} \frac{m_N}{(2\pi)^5(p_1 \cdot p_2)}
\sum_{\lambda_\gamma \lambda_p \lambda_\Sigma}
{|\cal M|}^2 \frac{d^3p_3}{2E_K} \frac{d^3p_4}{2E_{\pi}} \frac{M_{\Sigma}d^3p_5}{E_{\Sigma}} \nonumber \\
&&\times \delta^4(p_1+p_2-p_3-p_4-p_5),
\end{eqnarray}
where $\lambda_\gamma,\lambda_p,\lambda_\Sigma$ are the helicities
of the photon, proton and $\Sigma$, respectively.

For the one-pole case, the total amplitude can be represented by
\begin{eqnarray}
    {\cal M} = {\cal M}_K + {\cal M}_N + {\cal M}_{K^*} +
 {\cal M}_Y +  {\cal M}_C.
\end{eqnarray}
\begin{table}[htbp]
\caption{Values for the parameters taken from other references.}
\begin{tabular}{cccc}
\hline\hline
parameter & value & parameter & value \\
\hline
e & 0.303 & $\Lambda_M$ & 2.0GeV\cite{Nam:2015yoa} \\
$\kappa_N$ & 1.79\cite{Kim:2017nxg} & $g_{\gamma K K^*}$ & -0.254/GeV\cite{Kim:2017nxg} \\
$g_{K N \Lambda}$ & -13.4\cite{Stoks:1999bz} & $g_{K N \Sigma}$ & 4.09\cite{Stoks:1999bz} \\
$F_\pi$ & 92.2MeV\cite{Vysotsky:2015mks} & $g_{\Lambda^*K^*
N}$\footnote{Here we use  $g_{\Lambda^* K^* N}$ to denote the
$\Lambda(1405)\bar K^*N$ coupling constant for the one-pole case.} & 1.30\cite{Khemchandani:2011mf} \\
$g_{\Lambda^*_L K^* N}$ & 1.30\cite{Khemchandani:2011mf} & $g_{\Lambda^*_H K^* N}$ & 3.75\cite{Khemchandani:2011mf} \\
$\gamma_L$$^b$ & 0.85\cite{Nam:2003ch} & $\gamma_H$\footnote{Here we define $\gamma_{L/H}=g_{\Lambda^*_{L/H} KN}/g_{\Lambda^*_{L/H} \pi \Sigma}$.} & 2.37\cite{Nam:2003ch} \\
\hline\hline
\end{tabular}
\label{tab1}
\end{table}

Here the amplitude $\CM_Y$ represents the contribution from the
hyperon exchange diagrams(Fig.\ref{feynman}(b)). In principle, both
the $\Lambda$ and $\Sigma$ exchange amplitudes should be taken into
account explicitly. While due to the poor knowledge of the
$\Lambda^* Y\gamma$ coupling and the minor role of their
contributions in the present reaction, we take the
$\CM_Y=\CM_\Lambda$ and set $\mu_{\smsm{\Lambda^*\Lambda}}$ as a
free parameter to effectively take into account the sum of their
contributions. So in the following discussions, we will not
distinguish their individual contributions. To evaluate the
amplitudes, the parameters in the amplitudes, such as the coupling
constants, cutoff parameters and the parameters of the resonance,
need to be determined. In principle, all these parameters need to be
determined by fitting to the experimental data. To reduce the number
of free parameters, some parameters are fixed with the values
obtained in previous studies. As mentioned above, we list the values
of the parameters taken from other studies in Table \ref{tab1}. For
other parameters, we fit them to the data of the $\gamma p \to K^+
\pi\Sigma$ reaction. Now we have four free parameters, which are
three coupling constants($\mu_{\smsm{\Lambda^* \Lambda}}$,
$g_{\Lambda^* \bar K N}$, $g_c$) and one cutoff
parameter($\Lambda_B$). To determine these parameters, we fit them
to the recent data from the CLAS collaboration, which include the
angular distributions of the $K^+$, invariant mass spectrum of the
$\pi\Sigma$ and the polarization data. Note that at center of mass
energies larger than 2.3 GeV the $K^+$ angular distribution or the
$M_{\pi\Sigma}$ spectrum for the three charged channels of the
$\pi\Sigma$ system are similar to each other. Due to the relatively
large uncertainties of the data and for simplicity, we use the sum
data of the $\pi^+\Sigma^-$, $\pi^-\Sigma^+$ and $\pi^0\Sigma^0$
channels for the final analysis as in Ref.\cite{Nam:2015yoa}. For
later convenience, here we define $\gamma=g_{\Lambda^*
KN}/g_{\Lambda^* \pi \Sigma}$. In the fitting, we set $\gamma$ as
free parameter, and then the $g_{\Lambda^* KN}$ is determined by the
product of the $\gamma$ and $g_{\Lambda^* \pi \Sigma}$. By fitting
to the data, the free parameters are determined and presented in
Table \ref{tab2}. The fitting results for the total cross sections,
the angular distributions, the $\pi\Sigma$ invariant mass spectrum
and the $\Sigma$ polarization are shown by the solid line in Figs.
(\ref{cross_section})$-$(\ref{polarization}).
\begin{table}[htbp]
\caption{Fitted parameters for the one-pole case
($\chi^2/dof$=2.41).}
\begin{tabular}{cccc}
\hline\hline
parameter & value & parameter & value \\
\hline
$\Lambda_B$ & $2.00\pm0.05$ GeV & $\gamma$ & $2.36\pm0.02$ \\
$\mu_{\Lambda^*\Lambda}$ & $0.077\pm0.002$ & $g_c$ & $-8.57\pm0.12$ \\
\hline\hline
\end{tabular}
\label{tab2}
\end{table}

To explore the possible effects due to the two-pole structure of the
$\Lambda(1405)$,  we also need to discuss the formalism for the
two-pole case. For the two-pole case, we assume there are two I=0
resonances, i.e. $\Lambda^*_L$ and $\Lambda^*_H$, in the
$\Lambda(1405)$ region, and then the productions of both these two
resonances need to be considered in the full amplitude. Here the
subscripts $L$ and $H$ denote the states corresponding to the low-
and high- mass poles of the $\Lambda(1405)$ respectively. Since the
two resonances have same quantum numbers, the Feynman diagrams and
the structures of the amplitudes for the two resonances are
basically same and can be presented as the forms in Eqs.(18)-(22).
The new ingredients mainly come from the number of independent
amplitudes and the parameters in the amplitudes. The total amplitude
for the two-pole case can be written as
\begin{eqnarray}
{\cal M} &=& ({\cal M}^L_K + {\cal M}^L_N +  {\cal M}^L_{K^*} + {\cal M}^L_{Y}) + ({\cal M}^H_K  \nonumber \\
&& + {\cal M}^H_N +  {\cal M}^H_{K^*} + {\cal M}^H_{Y})e^{i\phi} +
{\cal M}_C e^{i\phi_c}
\end{eqnarray}
where $\phi$ and $\phi_c$ are introduced to describe the relative
phases among the amplitudes for the two resonances and the contact
term\footnote{It should be noted that a relative phase between the
resonance production amplitudes and the contact term can also be
introduced in the one-pole case. While, we find the final results
are not sensitive to this phase. So the introduction of this phase
in the one-pole case will not change the results presented below.}.
Some previous studies show that the relative phase between the two
resonance is about
$\pi$\cite{Nam:2015yoa,Schumacher:2013vma,Molina:2015uqp}.
Therefore, we adopt $\phi=\pi$ in this work. Since it is difficult
to constrain all the parameters by fitting the data of a single
reaction, we adopt the ChUM predictions for some of the parameters.
The values for these parameters are taken as: $g_{\Lambda^*_L K^*
N}=1.30$, $g_{\Lambda^*_H K^* N}=3.75$ \cite{Khemchandani:2011mf}
\footnote{At present, our knowledge of the coupling constants
$g_{\Lambda^*_{H/L} K^* N}$ are still rather limited. Here we adopt
one set of values of the $g_{\Lambda^*_{H/L} K^* N}$ predicted in
Ref.\cite{Khemchandani:2011mf} in the calculations. While, since the
$K^*$ exchange contribution only plays a minor role in this
reaction, if we adopt other predictions for these two coupling
constants in Ref.\cite{Khemchandani:2011mf} the results will not
change significantly.}, $\gamma_L=0.85$ and
$\gamma_H=2.37$\cite{Nam:2003ch}. Furthermore, for the same reason
as the one-pole case, we will also take into account the
contribution of hyperon exchanges by considering the effective
amplitudes $\CM_\Lambda^{L,H}$ with assuming
$\mu_{\smsm{\Lambda_L^*\Lambda}} = \mu_{\smsm{\Lambda_H^*\Lambda}}$.
Other parameters will be fixed by fitting the experiment data.

\begin{table}[htbp]
\caption{Fitted parameters for the two-pole case
($\chi^2/dof$=2.13).}
\begin{tabular}{cccc}
\hline\hline
parameter & value & parameter & value \\
\hline
$\Lambda_B$ & $1.94\pm0.09$ GeV & $\mu_{\Lambda^*_{\smsm{{L/H}\Lambda}}}$ & $0.069\pm0.006$ \\
$g_c$ & $8.35\pm0.16$ & $\phi_c$ & $-0.92\pm0.12$ \\
$M_{\Lambda^*_L}$ & $1357.9\pm1.1$ MeV & $M_{\Lambda^*_H}$ & $1425.1\pm3.4$ MeV \\
$g_{\Lambda^*_L \pi \Sigma}$ & $1.13\pm0.05$ & $g_{\Lambda^*_H \pi \Sigma}$ & $1.04\pm0.07$ \\
\hline\hline
\end{tabular}
\label{tab3}
\end{table}

\section*{RESULTS AND DISCUSSION}

\begin{figure}[htbp]
\begin{center}
\includegraphics[scale=0.46]{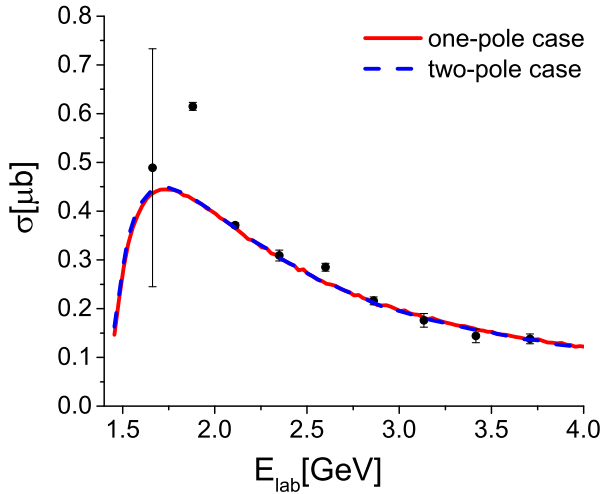}
\caption{Total cross sections for the $\gamma p \rightarrow K^+ \pi
\Sigma$ reaction as a function of the photon laboratory energy
$E_{lab}$. The red-solid and blue-dashed lines indicate the one-pole
and two-pole results, respectively. The data are taken from
Ref.~\cite{Moriya:2013hwg}.} \label{cross_section}
\end{center}
\end{figure}

With the help of the cernlib package MINUIT  and the formalisms
presented in the last section, the free parameters are fitted to the
experimental data from the CLAS collaboration\cite{Moriya:2013hwg}.
In Fig.~\ref{cross_section}, we show the total cross sections for
the $\gamma p \rightarrow K^+ \pi \Sigma$ reaction as a function of
the photon laboratory energy $E_{lab}$, where the solid and dashed
lines correspond to the results for the one-pole case and two-pole
case respectively. As discussed in Sec. I, only the data at the c.m.
energies(W) ranging from $2.3$ GeV to $2.8$ GeV are fitted in this
work. While, the comparison between our results and the experimental
data at $W<2.3$ GeV are also shown for completeness. The significant
discrepancy at the near threshold region can be attributed to the
ignoring of the $s$-channel nucleon resonance contributions. At
higher energies, their contributions are expected
small\cite{Kim:2017nxg}.

In Tables \ref{tab2} and \ref{tab3}, we present the fitted
parameters for the one-pole and two-pole cases. For the one-pole
case, the mass and width of the $\Lambda(1405)$ have been taken as
the values suggested by PDG. The coupling constant
$g_{\Lambda^*\pi\Sigma}$ can be determined through the decay width.
While, the coupling constants $g_{\Lambda^*\bar KN}$ will be
determined by fitting the experimental data. In literatures, the
values for these two coupling constants and their ratio have been
intensively studied(see Ref.\cite{Braun:1977wd,GellMann:1968rz} for
a detailed discussion) and our current knowledge about these
parameters still has large uncertainties. For example, the ratio of
$g_{\Lambda^*\bar KN}$ to $g_{\Lambda^*\pi\Sigma}$ may vary in a
range of 1.6-7.8\cite{Xie:2013wfa}. We find our fitting results of
these coupling constants and their ratio are consistent with
previous studies. This may give us some confidence about the
reliability of our model.  For the two-pole case, the masses of the
two resonances are set as free parameters. In the fitting, we find
that if no constraint is imposed the fitting will converge on some
solutions which is equivalent to the one-pole case. To pick out the
solution corresponding to the prediction of the ChUM, we have
demanded the masses of the two resonances should satisfy the
condition $M_{\Lambda^*_L}<1.4$ GeV and $M_{\Lambda^*_H}>1.4$
GeV\cite{Roca:2017wfo}. With this constraint condition, the obtained
masses of the two resonances are 1357.9 MeV and 1425.1 MeV,
respectively. Using the fitted values for the coupling constants
$g_{\Lambda^*_L\pi\Sigma}$ and $g_{\Lambda^*_H\pi\Sigma}$, the decay
width of the resonances can be obtained, and we get
$\Gamma_{\Lambda^*_L}=47$ MeV and $\Gamma_{\Lambda^*_H}=75$ MeV.

\begin{figure}[htbp]
\begin{center}
\includegraphics[scale=0.48]{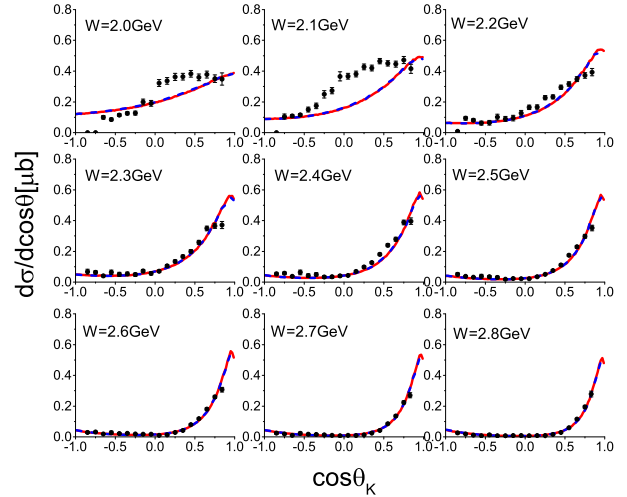}
\caption{The angular distributions of the $K^+$ in the center of
mass frame with $\theta_K$ being the angle between the $K^+$
momentum and the beam direction. The legends for the lines are the
same as those for the Fig.~\ref{cross_section}. The data are taken
from Ref.~\cite{Moriya:2013hwg}.} \label{angle_distribution}
\end{center}
\end{figure}

\begin{figure}[htbp]
\begin{center}
\includegraphics[scale=0.46]{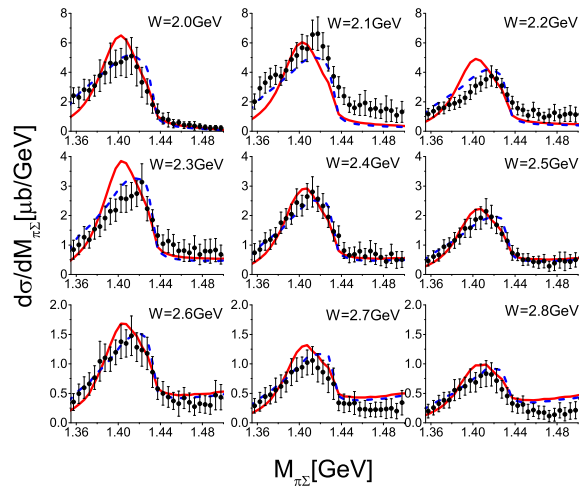}
\caption{The invariant mass spectrum of the final $\pi\Sigma$
system. The legends for the lines are the same as those for the
Fig.~\ref{cross_section}. The data are taken from
Ref.~\cite{Moriya:2013eb}.} \label{mass_spectrum}
\end{center}
\end{figure}

 The fitting results for the angular distribution and the
invariant mass spectrum are presented in
Figs.~\ref{angle_distribution} and \ref{mass_spectrum}. Our results
show that in our models both the one-pole and two-pole pictures can
give a good description of the data at the energies $W> 2.3$ GeV.
While, at the energies $W \leq 2.3$ GeV the two-pole picture gives a
better description of the invariant mass spectrum, even though only
the data with $W\geq 2.3$ GeV are considered in the fitting. For the
angular distributions, the results of the one-pole and two-pole
models overlap with each other, which shows that the angular
distribution is insensitive to the pole structures of the
$\Lambda(1405)$. In the angular distributions, the $t$-channel $K$
and $K^*$ meson exchanges are responsible for the enhancement at
forward angles, and the $u$-channel hyperon exchange results in the
slight enhancement at backward angles.

After a brief discussion of the results of the total cross sections,
angular distributions and invariant mass spectrums, let us come to
the $\Lambda(1405)$ polarization in this reaction. In
Ref.\cite{Moriya:2014kpv}, the CLAS collaboration has explored the
polarization of the $\Lambda(1405)$  in the reaction $\gamma p
\rightarrow K^+ \Lambda(1405)$ at $0.6<cos\theta^{c.m.}_{K^+}<0.9$
in the energy range $2.55 < W < 2.85$ GeV. Based on some general
arguments, they concluded that using an unpolarized beam and target
the polarization of the $\Lambda(1405)$ can only happen in the
direction out of the production plane, and in the process
$\Lambda(1405) \rightarrow \pi + \Sigma$ the polarization of
$\Sigma$ is exactly the same as the polarization of the
$\Lambda(1405)$ in the $\Lambda(1405)$ rest frame regardless of
decay angle if the $\Lambda(1405)$ is a s-wave state. Based on their
measurement and analysis, they found the experimental data supported
the s-wave nature of the $\Lambda(1405)$, which is the first
experimental determination of the $J^P$ quantum numbers of the
$\Lambda(1405)$. Since the produced $\Lambda(1405)$ is polarized in
the present reaction, it will then be interesting to discuss the
mechanism for its polarization. Furthermore, it is also interesting
to discuss the possible different predictions of the $\Lambda(1405)$
polarization in the one-pole or two-pole pictures of the
$\Lambda(1405)$. To our best knowledge, such questions are still not
discussed in previous
works\cite{Nakamura:2013boa,Roca:2013av,Nam:2015yoa,Oset:2015ksa}.

\begin{figure}[htbp]
\begin{center}
\includegraphics[scale=0.45]{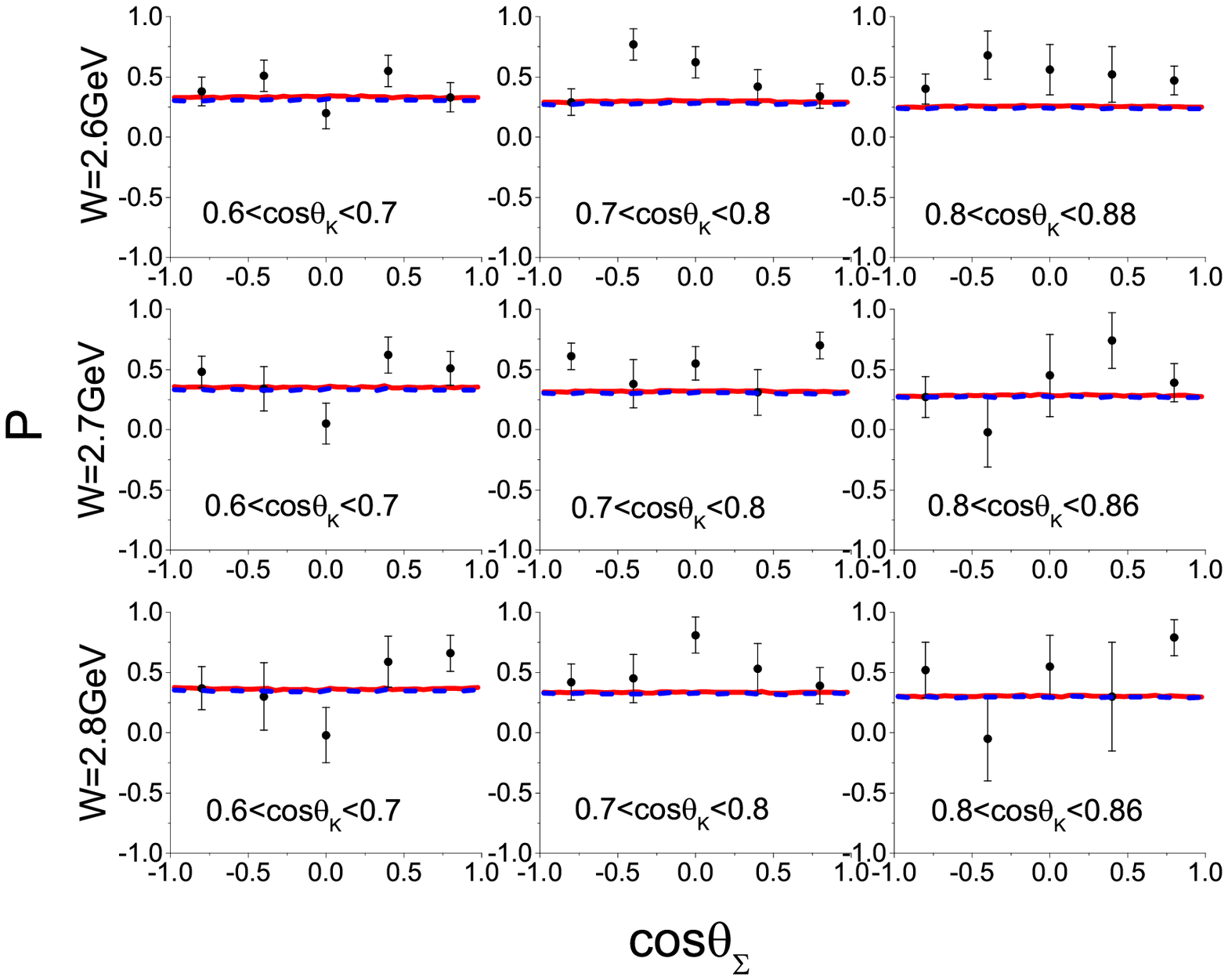}
\caption{The polarization of the final $\Sigma$ as a function of
$cos\theta_{\Sigma}$ in the rest frame of $\pi\Sigma$ for the chosen
kinematic bins. The polarization axis is taken along
$\vec{p}_\gamma\times \vec{p}_{K^+}/|\vec{p_\gamma}\times
\vec{p}_{K^+}|$ and the $\theta_\Sigma$ is defined as the angle
between the $\Sigma$ momentum and the polarization axis. The legends
for the lines are the same as those for the
Fig.~\ref{cross_section}. The data are taken from
Ref.~\cite{Moriya:2014kpv}.} \label{polarization}
\end{center}
\end{figure}

First, we tried to reproduce the polarization data by only
considering the $\Lambda(1405)$ production
amplitudes(Fig.\ref{feynman}a-\ref{feynman}c). In such a model, we
can describe the angular distribution and the invariant mass
spectrum well as in Ref.\cite{Kim:2017nxg}. However, the
polarization data can not be reproduced. We then introduce a contact
term(Fig.\ref{feynman}d) in the present model. After including the
contact term contribution, now we can interpret the $\Sigma$
polarization data well. As shown in Fig.~\ref{polarization}, the
polarization of $\Sigma$ is almost flat in the $\Lambda^*$ rest
frame with the polarization axis being along the direction out of
the production plane, which is consistent with the $s$-wave nature
of the $\Lambda(1405)$. As can be seen from the figures, our result
agrees well with the current experiment data. We also find that both
the one-pole and two-pole pictures can give a good description of
the polarization data. It is then interesting to ask whether one can
find some observable to distinguish the one- or two-pole models in a
single reaction. As we know, the distinct feature of the two-pole
picture is that the two poles are near and may have different
coupling strengths to $\pi\Sigma$ and $\bar K N$ channels. It is
then natural to expect that the strengths of the contributions of
the two poles are different and their relative roles may change as
the invariant mass $M_{\pi\Sigma}$ crosses the $\Lambda(1405)$
region. Since the polarization observables are sensitive to the
interference term and thus the change of the relative roles of the
two poles, it is possible that the polarization of
$\Lambda(1405)$(or more accurately, the polarization of the
$\Sigma$) may have a quite different dependence on the invariant
mass $M_{\pi\Sigma}$ in the one-pole and two-pole pictures.

\begin{figure}[htbp]
\begin{center}
\includegraphics[scale=0.45]{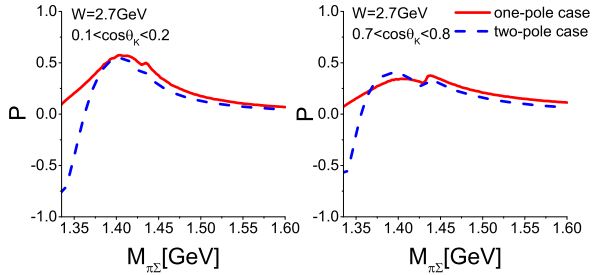}
\caption{The polarization of $\Sigma$ as a function of $M_{\pi
\Sigma}$ at the c.m. energy $W=2.7$ GeV for
$0.1<cos\theta^{c.m.}_{K^+}<0.2$ and
$0.7<cos\theta^{c.m.}_{K^+}<0.8$.} \label{polarization_with_mass}
\end{center}
\end{figure}

\begin{figure}[htbp]
\begin{center}
\includegraphics[scale=0.45]{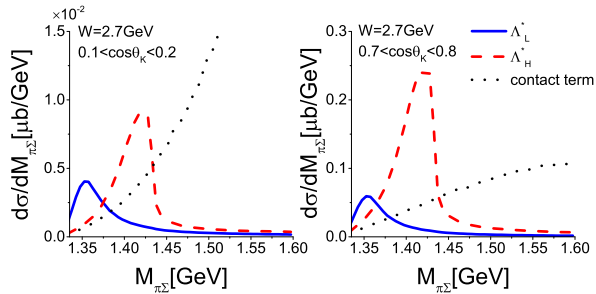}
\caption{The contributions of the $\Lambda^*_L$, $\Lambda^*_H$ and
contact term in the $M_{\pi\Sigma}$ spectrum at the c.m. energy
$W=2.7$ GeV for $0.1<cos\theta^{c.m.}_{K^+}<0.2$ and
$0.7<cos\theta^{c.m.}_{K^+}<0.8$.} \label{mass_with_angle}
\end{center}
\end{figure}

In Fig.\ref{polarization_with_mass}, we study the polarization of
the final $\Sigma$ versus $M_{\pi\Sigma}$ at some angle
bins\footnote{It is worth noting that the $\Sigma(1385)$ and the
$K^*(892)$ also contribute in this reaction. For the $\Sigma(1385)$,
due to its small coupling to the $\pi\Sigma$ channel and its weak
interferences with other contributions, it only plays a minor role
here. We have checked that in our model the inclusion of the
$\Sigma(1385)$'s contribution does not significantly change the
results presented below. The $K^*(892)$'s contribution is not
considered because it also plays a minor role in this
reaction\cite{Moriya:2010yya}. Furthermore, at the energies
considered in this work the bands of the $K^*(892)$ and the
$\Lambda(1405)$ are well separated in the Dalitz
plot\cite{Moriya:2014kpv}. So it is possible to eliminate its
contribution by a cut on the invariant mass of the $K\pi$ system.
After a cut on the $K^*(892)$'s contribution, it will be safe to
ignore its contribution and the effects due to its finite width on
the results presented below.}. It is found the polarization of the
$\Sigma$ indeed shows different patterns in the $\Lambda(1405)$
region for the two pictures. Through a more detailed study, we find
the $\Sigma$ polarization originates from the interference between
the $\Lambda(1405)$ production amplitude and the contact term, and
the polarization is sensitive to their relative phases. If the
two-pole picture is correct, the relative roles of the two poles may
change in the $\Lambda(1405)$ region, which may result in a strong
dependence of the $\Sigma$ polarization on the $M_{\pi\Sigma}$. In
Fig.\ref{mass_with_angle} we present the contributions of the
$\Lambda_L^*$, the $\Lambda_H^*$ and the contact term in the
invariant mass spectrum at the same angle bins as in
Fig.\ref{polarization_with_mass}. It is found that the $\Sigma$
polarization may show strong dependence on the $M_{\pi\Sigma}$ at
the place where the contributions of the two resonances are
comparable. Here we need to note that the pattern of the $\Sigma$
polarization shown in Fig.\ref{polarization_with_mass} is dependent
on the relative phase between the two resonances, which is set as
$\phi=\pi$ in this work. By adopting a different value for $\phi$,
the pattern of the $\Sigma$ polarization will change. However, the
strong dependence of the $\Sigma$ polarization on the
$M_{\pi\Sigma}$ remains at the place where the two resonances have
comparable contributions\footnote{In this work, we have also tried
to set $\phi$ as 0, $\pi/2$ or $3\pi/2$. In these cases, the strong
dependence of the $\Sigma$ polarization on the $M_{\pi\Sigma}$
remains. However, we can not get a good description of the invariant
mass spectrum.}. Therefore, we expect the measurement of the
$\Sigma$ polarization versus the invariant mass $M_{\pi\Sigma}$ in
the $\gamma p\to K^+ \Lambda(1405)$ reaction may offer the chance to
verify the pole structure of the $\Lambda(1405)$.

Based on the above discussions, we conclude that our model results
show the measurement of the $\Sigma$ polarization versus
$M_{\pi\Sigma}$ may verify the two-pole picture of the
$\Lambda(1405)$ predicted by the ChUM. Until now, there is still no
final conclusion about this issue. So it is important to find some
new way to distinguish the two pictures of the $\Lambda(1405)$.
Although the results presented above may have model dependence, the
argument about the different dependence of the $\Sigma$ polarization
on the $M_{\pi\Sigma}$ in the two pictures may also hold for other
models. Since the polarization observable is more sensitive to the
interference terms among the amplitudes, we expect the polarization
observable may offer more clues about the pole structure of the
$\Lambda(1405)$. Furthermore, we also expect that a similar
conclusion can be made for other $\Lambda(1405)$ production
processes, where the produced $\Lambda(1405)$ is polarized.
\section*{SUMMARY}
In this work, we investigate the polarization of the $\Lambda(1405)$
in the $\gamma p \rightarrow K^+ \pi \Sigma$ reaction. We consider
the contributions from the t-channel $K/K^*$ exchanges, the
u-channel hyperon exchange and a contact term. In our model, the
contact term is necessary for interpreting the $\Lambda(1405)$
polarization. In addition, we also find that although both the
one-pole and two-pole models can give a good description of the
angular distribution and the invariant mass spectrum data, they give
distinct predictions for the polarization of the final $\Sigma$
versus the $M_{\pi\Sigma}$. Thus the measurement of the dependence
of the $\Sigma$ polarization on the $M_{\pi\Sigma}$ can offer
valuable information about the pole structure of the
$\Lambda(1405)$. To make this measurement, a large statistics of
data will be needed. We hope such a measurement can be done in the
future.

\begin{acknowledgements}

We acknowledge the support from the National Natural Science
Foundation of China under Grants No.U1832160 and No.11375137, the
Natural Science Foundation of Shaanxi Province under Grant
No.2019JM-025, and the Fundamental Research Funds for the Central
Universities.
\end{acknowledgements}

\end{document}